\documentclass{caosp310}

\usepackage{amsthm}

\usepackage[a-2u]{pdfx}

\usepackage{rotating} % For rotating tables
\usepackage{graphicx} % For \resizebox
\usepackage{makecell} % For multi-line headers
\usepackage{pdflscape} % Include this in your preamble
\usepackage{longtable} % For tables spanning multiple pages
\usepackage{lscape} % For rotating the page landscape
\usepackage{float}    % For the H placement option

\usepackage{hyperref}    % For hyperlinks (optional, but often useful)

\usepackage[utf8]{inputenc}

\usepackage{lmodern}
\usepackage{amssymb}
\usepackage{subfigure}
\usepackage{amsmath}        % extensions for typesetting of math
\usepackage{amsfonts}       % math fonts
\usepackage{amsthm}         % theorems, definitions, etc.
\usepackage{bbding}         % various symbols (squares, asterisks, scissors, ...)
\usepackage{bm}             % boldface symbols (\bm)
\usepackage{fancyvrb}       % improved verbatim environment
\usepackage{natbib}         % citation style AUTHOR (YEAR), or AUTHOR [NUMBER]
\usepackage{dcolumn}        % improved alignment of table columns
\usepackage{booktabs}       % improved horizontal lines in tables
\usepackage{paralist}       % improved enumerate and itemize
\PassOptionsToPackage{usenames}{xcolor}
\usepackage{xcolor}
\usepackage[utf8]{inputenc}   % For encoding
\usepackage[T1]{fontenc}      % For proper character representation in PDF
\usepackage[nottoc]{tocbibind} % makes sure that bibliography and the lists
\usepackage{dcolumn}        % improved alignment of table columns
\usepackage{booktabs}       % improved horizontal lines in tables
\usepackage{txfonts}
\usepackage{natbib}
\articleNo{}
\pubyear{2025}
\volume{55}
\volnumber{1}
\firstpage{60}
\received{March 23, 2025}
\accepted{March 31, 2025}

\def\BibTeX{{\rm B\kern-.05em{\sc i\kern-.025em b}\kern-.08em
             T\kern-.1667em\lower.7ex\hbox{E}\kern-.125emX}}

\begin{document}

\hauthor{T.\,Ramezani {\it et al.}}

   \title{Ultraviolet Photometry and Reddening Estimation of 105 Galactic Open Clusters}

   \author{T. Ramezani\inst{1}\orcid{0009-0005-6112-479X}
          \and
          E. Paunzen\inst{1}\orcid{0000-0002-3304-5200}
          \and
          A. Gorodilov\inst{1}\orcid{0009-0004-7058-0802}
          \and
          O. I. Pintado\inst{2}\orcid{0000-0003-4650-950X}
          }

   \institute{Department of Theoretical Physics and Astrophysics, Masaryk University, Kotl\'a\v{r}sk\'a 2, 611\,37 Brno, Czechia  \\ \email{epaunzen@physics.muni.cz} 
    \and Consejo Nacional de Investigaciones Cient{\'i}ficas y T{\'e}cnicas, Argentina \\ }    

\date{Received: March 23, 2025; Accepted: March 31, 2025}
\maketitle

\begin{abstract}
This paper focuses on observing unstudied Galactic open clusters 
in the Ultraviolet (UV) wavelength range and analyzing their photometric data. 
The Gaia Data Release 3 (DR3) enables us to precisely study known Galactic open clusters.  
We conducted observations using the 1.54-meter Danish Telescope (DK1.54) in Chile 
and the 2.15-meter telescope at the Complejo Astron{\'o}mico El Leoncito (CASLEO) in 
Argentina, employing UV filters.  
Furthermore, we have collected available photometric and astrometric data for our observed clusters. 
We aim to estimate the reddening of Galactic open clusters using UV photometry.
We applied isochrone fitting to determine the reddening of the clusters
using well-known members.
As a final result, we present the reddening values of 105 Galactic open clusters in the UV, 
as determined by our photometry.
\keywords{Open Clusters, Gaia, Ultraviolet, Photometry}
\end{abstract}

\section{Introduction}

Star clusters, particularly open clusters, serve as essential laboratories for understanding 
stellar formation, evolution, and dynamics. Since stars within a cluster share a common 
origin, distance, and age, their study provides crucial insight into the broader processes 
of stellar evolution \citep{2020SSRv..216...64K}. By analyzing star clusters, we can refine our models of stellar 
formation and lifecycles, such as those governed by mass, composition, and environment. 
Studying these clusters in the Ultraviolet (UV) wavelengths is particularly important 
because UV observations are sensitive to young, hot, and massive stars \citep{galaxies8030060}. These stars 
emit most of their light in the UV, making this wavelength regime critical for 
understanding the early stages of stellar evolution and the Interstellar Medium (ISM). 
One of the key challenges in studying star clusters is interstellar reddening \citep[absorption 
or extinction;][]{2003A&A...397..191P}. This effect is caused by dust grains in the ISM that scatter and absorb 
starlight, causing the light from stars to shift toward redder wavelengths \citep{1989ApJ...345..245C}. The amount 
of reddening is a function of the dust column density and the object's location within 
the Galaxy. Correcting for this reddening is essential for accurately determining 
stellar properties such as temperature and luminosity. Extinction models, such as 
the \citet{Fitzpatrick1999} model, are commonly used to correct for these effects. 
These models help to remove the distortion introduced by the ISM and allow us to 
recover the intrinsic colours and magnitudes of stars.

Unfortunately, recent UV observations of star clusters are very scarce. Only a few
Well-studied ones are also analysed in the UV \citep{2018MNRAS.481..226S}. The
Ultra Violet Imaging Telescope (UVIT) consortium presented some results 
for well-known star clusters \citep{2021MNRAS.503..236J}.

In this paper, we present the status of our project to observe Galactic open clusters
to study their extinction characteristics using our own observed Johnson $U$ and 
archival Gaia $BP$, $RP$, and $G$ photometry.

\section{Target selection and observations}

For the first case study of our project, we selected 105 southern star 
clusters from the list of \citet{Hunt2023} which are not too extended on the sky.
This catalogue contains the parameters (age, reddening, and distance) and positional 
information of 7\,167 star clusters,
including moving groups, 
with over 700 newly discovered high-confidence clusters. They used the widely applied
Hierarchical Density-Based Spatial Clustering of Applications with Noise (HDBSCAN) 
algorithm \citep{McInnes2017}.

We conducted observations using the 1.54-meter Danish Telescope (DK1.54) in Chile 
and the 2.15-meter telescope at the Complejo Astron{\'o}mico El Leoncito (CASLEO) in 
Argentina, employing UV filters.

The DK1.54 was equipped with the Danish Faint Object Spectrograph and Camera (DFOSC)
using a 2k × 2k thinned Loral CCD chip with a field of view (FOV) of 12×12 arc minutes. 
The 2.15-meter Jorge Sahade reflector at CASLEO was using a 2k × 2k
Roper Scientific Versarray 2048B camera with a FOV of about nine arc minutes.

The exposure time for each cluster is 300 seconds, with at least two observations for each cluster. 

\section{Data Reduction}

The basic CCD reductions (bias-subtraction, dark
correction, and flat-fielding) were performed with standard IRAF v2.17 routines. 
We removed cosmic rays whenever needed before 
calculating the Point Spread Function (PSF) and the instrumental magnitude for all stars. 

As the next step, we matched the observed fields using the 
Atlas of Large-Area Digital Image Navigation (Aladin). For this, we used the IRAF task
``xy2sky'' to transform pixel coordinates (x, y) into the celestial coordinates (RA, DEC).
After a rotation of 270 degrees, we could compare them with external catalogues. 

For our purposes, we used the Gaia DR3 \citep{2023A&A...674A..32B} and its photometry
($BP$, $RP$, and $G$ magnitudes) for the matching process. This was done by a newly developed 
pipeline software\footnote{https://github.com/PoruchikRzhevsky/Match-pipeline} based 
on the Match Program\footnote{http://spiff.rit.edu/match/}, which uses the 
FOCAS algorithm \citep{Valdes1995}. The script takes the Gaia data and our objects in .CSV 
format as input data. 
Loading confirmed cluster members representing main sequence stars 
into the script is also possible. We used \citet{Hunt2023} catalogue, which is based 
on Gaia DR3 data, to define members. We then filtered the matched stars for cluster members. 

\begin{figure}[!t]
  \centering
    \subfigure[]{\includegraphics[width=0.48\textwidth]{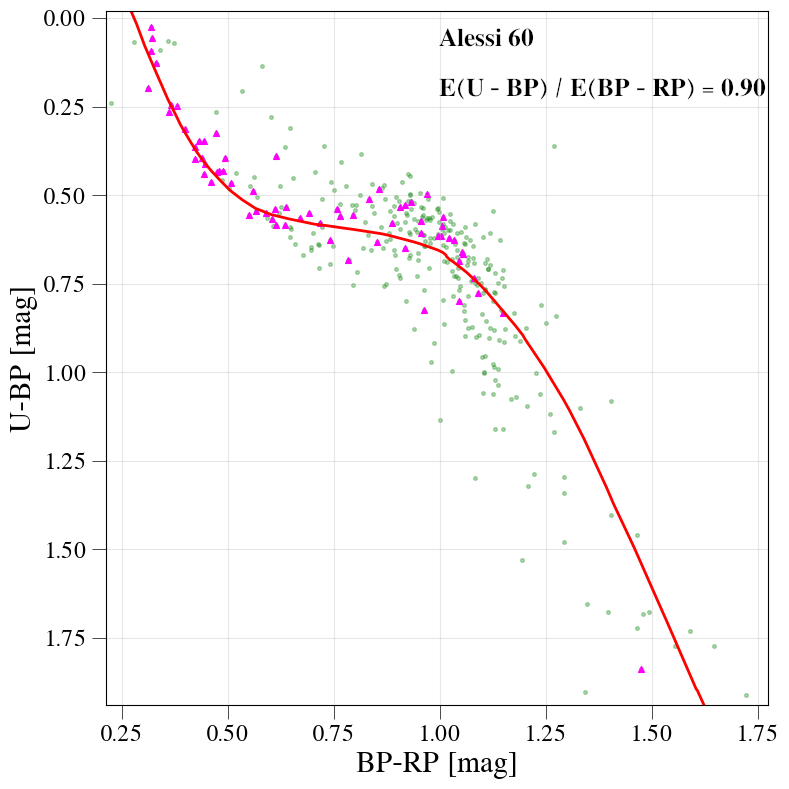}} 
    \subfigure[]{\includegraphics[width=0.48\textwidth]{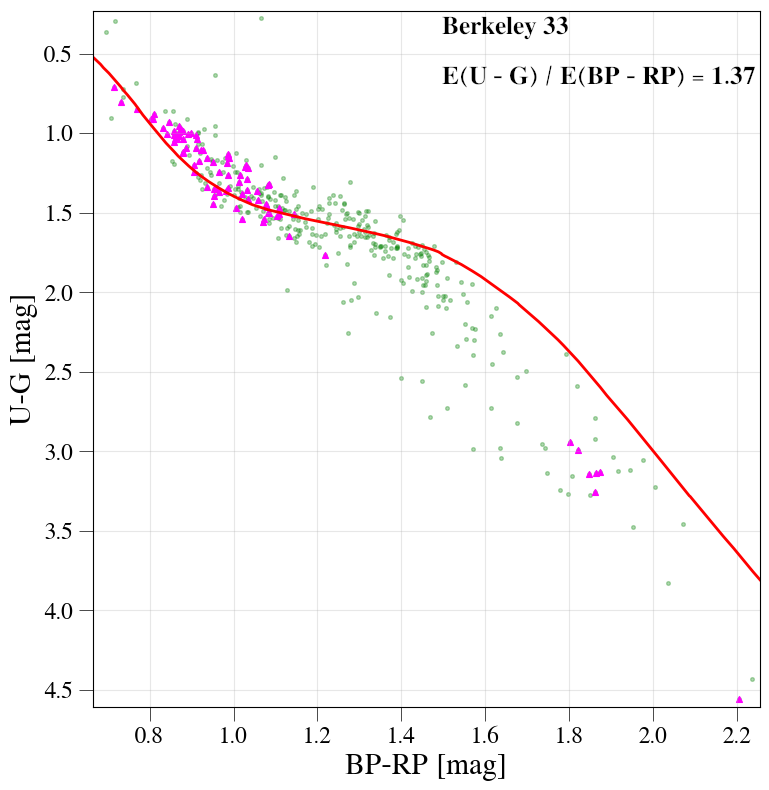}} \\
    \subfigure[]{\includegraphics[width=0.48\textwidth]{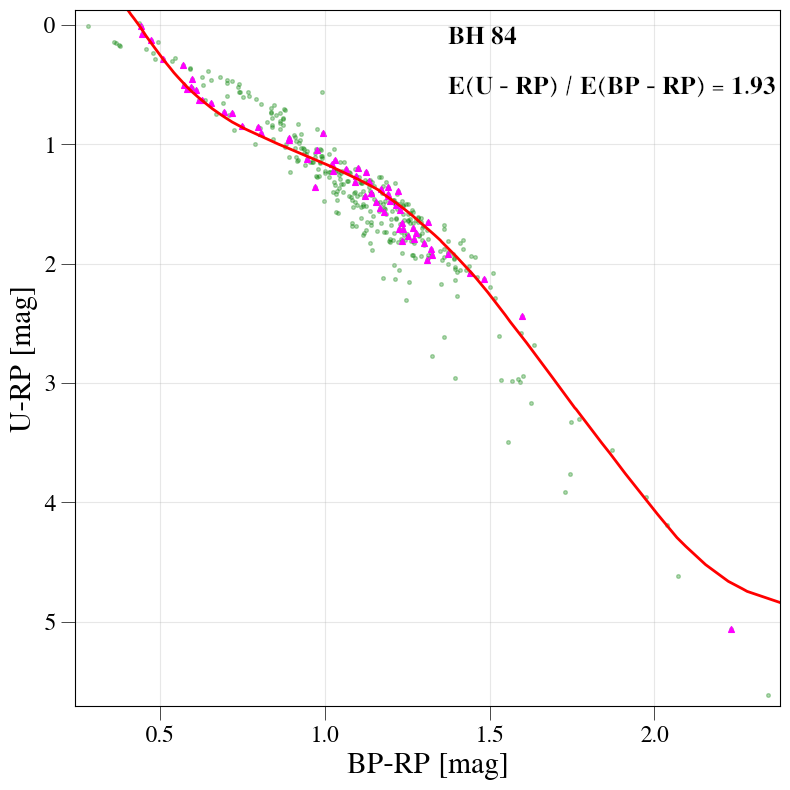}}
    \subfigure[]{\includegraphics[width=0.48\textwidth]{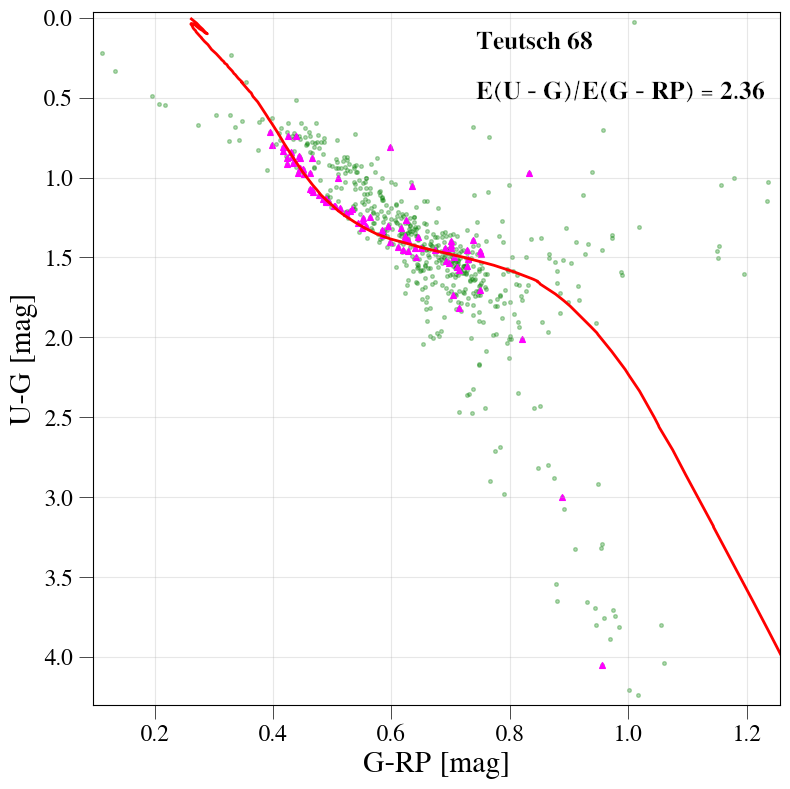}}
  \caption{Colour$-$Colour diagrams of different clusters and filter combinations.
  The standard lines are taken from \citet{Bressan2012}. The green dots are matched non-members
  and pink triangles are members from \citet{Hunt2023}.}
  \label{fig:sub1}
\end{figure}

Instrumental magnitudes are typically uncalibrated values from the instrument, which depend 
on the specific setup (filters, detectors, and exposure times).
The Gaia DR3 provides flux-calibrated low-resolution spectrophotometry (BP/RP spectra) 
for about 200 million sources in the wavelength range from
330 to 1050\,nm \citep{2023A&A...674A..33G}.
From these spectra, synthetic photometry can be derived for any passband. 
Existing observations can be reproduced within a few per cent over a wide range of 
magnitudes and colour for wide and medium bands, such as Johnson $U$, and with 
up to millimag accuracy 
when synthetic photometry is standardised for external sources. We first checked the
available standard $U$ photometry using The General Catalogue of Photometric Data\footnote{https://gcpd.physics.muni.cz/}. No offsets or systematics were detected.

As the last step, we transformed our instrumental magnitudes to standard ones as described
by \citet{2005ARA&A..43..293B}.

\section{Estimating the reddening values}

We generated colour-colour diagrams from Colour-Magnitude Diagrams (CMDs), 
including $U$ photometry and fitted the filtered stars to find their extinction in 
each colour from the main sequence standard line for the different colour combinations. 
The standard lines are taken from the Padova database of stellar evolutionary tracks 
and isochrones \citep{Bressan2012}.

Fitting is done manually, using a GUI interface that allows 
the standard main sequence to be shifted along two axes corresponding 
to the combination of colours selected. The script calculates the errors
for a confidence interval of 99.73\%.
A tutorial on the use of pipelines is available in the README.md file. 

We get the reddening from $U$ in the independent colour-colour diagram 
of age, distance, and metallicity. We estimate the reddening for all 
our observed clusters in different passbands. 

Figure \ref{fig:sub1} shows the Colour$-$Colour diagrams of different clusters and 
filter combinations. The green dots are matched non-members, and the pink triangles 
are members from \citet{Hunt2023}.

In Table \ref{table_values}, we present the name of the clusters, their coordinates 
in RA and DEC, different extinction ratios, uncertainties of each extinction's ratios, 
mean value of extinctions, the distance and log age from the
\citet{Hunt2023} catalogue. 
Only cluster BH 140 does not have log age in their catalogue. 
In its complete form, this table is only available at the CDS or upon request.

\begin{table*}
   \centering  
   \caption{Coordinates, reddening values, and uncertainties of clusters.  In its complete form, this table is only available at the CDS or upon request. The first page is printed here for guidance regarding its form and content. The columns denote: (1) Clusters' name. (2) Right ascension (J2000; Gaia DR3). (3) Declination (J2000; GaiaDR3). (4) $E(U-BP)/E(BP-RP)$. (5) $E(U-G)/E(BP-RP)$. (6) $E(U-RP)/E(BP-RP)$. (7) $E(U-G)/E(G-RP)$. (8) $E(U-G)/E(BP-G)$. (9) Uncertainty of $E(U-BP)/E(BP-RP)$. (10) Uncertainty of $E(U-G)/E(BP-RP)$. (11) Uncertainty of $E(U-RP)/E(BP-RP)$. (12) Uncertainty of $E(U-G)/E(G-RP)$. (13) Uncertainty of $E(U-G)/E(BP-G)$. (14) Mean extinction. (15) Distance (kpc). (16) $\log t$. }
   \resizebox{\textwidth}{!}{ % Resize to fit within page width
   \Large
   \begin{tabular}{llcccccccccccccccc}
   \hline\hline
   (1) & (2) & (3) & (4) & (5) & (6) & (7) & (8) & (9) & (10) & (11) & (12) & (13) & (14) & (15) & (16) \\
   \hline
   \large Alessi\_17 &\large 113.853 &\large -15.092 &\large 1.08 &\large 1.36 &\large 2.01 &\large 2.08 &\large 3.78 &\large 0.28 &\large 0.11 &\large 0.17 &\large 0.76 &\large 0.41 &\large 2.06 &\large 3.92 &\large 8.38 \\
   % \hline
   \large Alessi\_60 &\large 105.615 &\large -1.120 &\large 0.90	&\large 1.29 &\large 1.90 &\large 2.03 &\large  2.09 &\large 0.14 &\large 0.09 &\large 0.08 &\large 0.22 &\large  0.20 &\large 1.64 &\large 2.64 &\large 8.68  \\
   % \hline
   \large Berkeley\_33 &\large 104.454 &\large -13.226 &\large 0.98 &\large 1.37 &\large 1.98 &\large 2.24 &\large 3.45 &\large 0.08 &\large 0.10 &\large 0.08 &\large 0.29 &\large 0.16 &\large 2.00 &\large 4.73	&\large 8.68 \\
    % \hline
    \large BH\_72 &\large 142.843 &\large -53.041 &\large 1.30 &\large 1.68 &\large 2.30 &\large 2.18 &\large 2.98 &\large 0.55 &\large 0.56 &\large 0.57 &\large 0.80 &\large 0.88 &\large 2.09 &\large 4.51 &\large 8.23  \\
    % \hline
    \large BH\_84 &\large 150.334 &\large -58.217 &\large 0.92 &\large 1.33 &\large 1.93 &\large 2.21 &\large 3.22 &\large 0.12 &\large 0.13 &\large 0.13 &\large 0.34 &\large 0.35 &\large 1.92 &\large  3.80 &\large	8.21  \\
    % \hline
    \large BH\_87 &\large 151.162 &\large -55.377 &\large 1.13 &\large 1.54 &\large 2.13 &\large 2.52 &\large 3.71 &\large 0.11 &\large 0.10 &\large 0.08 &\large 0.21 &\large 0.27 &\large 2.21 &\large 2.18 &\large 8.15  \\
    % \hline
    \large BH\_111 &\large 167.317 &\large -63.830 &\large 0.91 &\large 1.30 &\large 1.91 &\large 2.15 &\large 3.32 &\large 0.23 &\large 0.24 &\large 0.22 &\large 0.51 &\large 0.53 &\large 1.92 & \large 2.46 &\large 8.43  \\
    % \hline   
    \large BH\_132 &\large 186.725 &\large -64.065 &\large 0.85 &\large 1.32 &\large 1.85 &\large 1.93 &\large 3.45 &\large 0.31 &\large 0.32 &\large 0.32 &\large 0.46 &\large 0.32 &\large 1.88 &\large 2.47 &\large 7.98 \\
    % \hline
    \large BH\_140 &\large 193.454 &\large -67.182 &\large 1.16 &\large 1.55 &\large 2.17 &\large 2.84 &\large 3.44 &\large 0.06 &\large 0.07 &\large 0.07 & \large 0.38 &\large 0.39 &\large 2.23 &\large 4.60 &   \\
    % \hline
    \large CWNU\_95 &\large 223.299 &\large -54.108 &\large 0.59 &\large 1.00 &\large 1.63 &\large 1.63 &\large 2.60 &\large 0.18 &\large 0.17 &\large 0.18 &\large 0.68 &\large 0.74 &\large 1.49 &\large 1.05 &\large 7.37 \\   
    % \hline 
    \large CWNU\_1733 &\large 114.240 &\large -26.321 &\large 0.39 &\large 0.89 &\large 1.45 &\large 1.51 &\large 2.11 &\large 0.27 &\large 0.32 &\large 0.28 & \large 0.71 &\large 0.80 &\large 1.27 &\large 1.73 &\large 8.30 \\
   % \hline
    \large Czernik\_29 &\large 112.095 &\large -15.399 &\large 1.08 &\large 1.41 &\large 2.08 &\large 2.44 &\large  3.66 &\large 0.09 &\large 0.10 &\large 0.07 &\large 0.32 &\large 0.21 &\large 2.13 &\large 3.51 &\large 8.39 \\
   \hline
   \hline  
   \label{table_values}
    \end{tabular}}
\end{table*}

\begin{figure}[!t]
  \centering
    \subfigure[]{\includegraphics[width=0.98\textwidth]{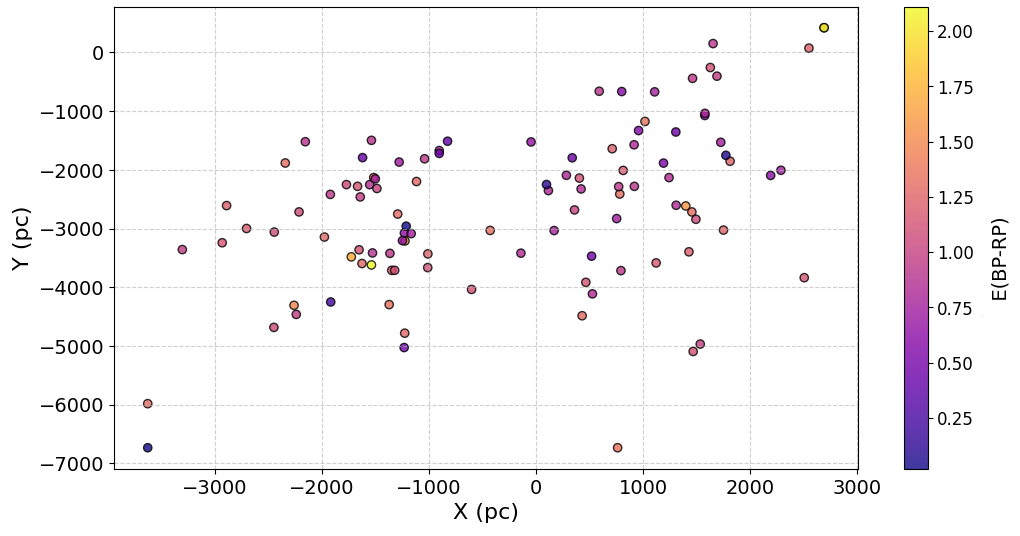}} \\
    \subfigure[]{\includegraphics[width=0.98\textwidth]{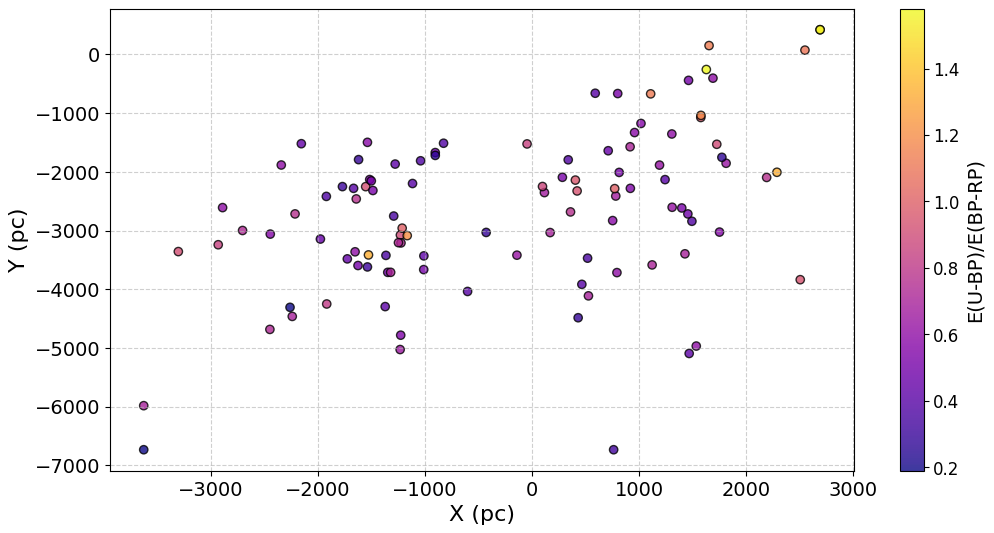}} 
  \caption{The upper panel shows the extinction in $(BP - RP)$. The lower panel shows the
  reddening ratio $E(U - BP)/E(BP - RP)$ for the
  105 observed clusters in the [X, Y] coordinate system around the Sun at [0, 0], respectively.}
  \label{fig:sub2}
\end{figure}

\section{Analysis}

Let us recall that stars in galaxies are born in molecular clouds (MCs). The gravitational 
collapse of the dense regions in the ISM of galaxies is an important mechanism 
to form star clusters \citep{Keilmann2024}. Dust is made up of heavy elements resulting from star 
nuclear burning. Dust grains are formed by reprocessing these heavy elements in the interstellar 
medium after they are expelled from stars by winds and explosions \citep{Draine2003}. 

Dust (and gas) significantly affects light propagation, scattering, and absorbing photons from 
UV to infrared wavelengths, leading to extinction and reddening effects in astronomical 
observations \citep{Schlafly2011}. There are two different effects of extinction: 1) 
From the molecular cloud that the star cluster is born from; 2) From the ISM between the observer 
and star cluster. 

Dust is not symmetrically distributed throughout space because of supernova explosions, 
stellar winds, magnetic fields, and star formation variations. Another complication is the 
composition of the ISM, which directly affects the reddening law, i.e.  
the variation of extinction with wavelength. Therefore, the reddening law and the amount of extinction
depend on the line of sight. It is well known that the Galactic disk has regions with an extinction
as high as five magnitudes per kpc \citep{1980A&AS...42..251N}. 

Expressing the extinction law is not unique in the literature; it has been common practice to use the 
ratios of two colours, for example, $E(U - B)/E(B - V)$. Using $A(V)$ as the reference extinction 
in the visual is arbitrary \citep{1989ApJ...345..245C}. 

In Fig. \ref{fig:sub2}, we present the results of our extinction estimation in [X, Y] coordinates.
The X coordinate is in the direction of the Galactic centre, whereas Y is in the direction of the 
disk rotation. The Sun is located at [0, 0], respectively. Because all studied star clusters
are located in the Galactic disk, we do not consider the third coordinate [Z] for our analysis.

With very few exceptions, the extinction is increasing for more distant clusters
(Fig. \ref{fig:sub2}, upper panel). This is what the current models predict, and it lends confidence in our
fitting procedure.

The distribution of the reddening ratio (Fig. \ref{fig:sub2}, lower panel) shows exciting features.
For example, there is a continuous transition in the direction [--3000, --3000], some sudden
changes on small scales [--1000, --3000], and no changes in certain lines of sight. It proves the capability to
trace the ISM with extinction estimates of star clusters.

\section{Conclusions and Outlook}

It is well known that photometric observations in the UV region help to describe the extinction law
and the total absorption. Unfortunately, because of the inefficiency of modern CCD detectors, such 
observations are not very common any more. On the other hand, the number of star clusters is
constantly increasing, caused by the excellent astrometric data of the Gaia satellite mission.

Here, we present our first case study using the synergy of ground-based UV observations and 
Gaia $BP$, $RP$, and $G$ photometry to analyse star clusters. High-precision PSF photometry using published
membership probabilities and a matching routine resulted in a unique data set of mainly unstudied 
open clusters.

Fitting standard main sequences, which are independent of age, metallicity and distance, allowed
us to get absorption values and reddening laws toward 105 open clusters. Detailed maps show that
our current models could be proven and have the potential to study the characteristics of the ISM in
more detail.

As the next step, we plan to observe more Galactic open clusters and generate a reddening law map
of the visible regions around the Sun. Another approach we will follow is the synthetic Gaia 
flux-calibrated low-resolution spectrophotometry (BP/RP spectra), which allows for a synthesis of photometry.

\begin{acknowledgements}
This work was supported by the grants GA{\v C}R 23-07605S, \\
MUNI/A/1077/2022, and MUNI/A/1419/2023. 
We want to thank doc. RNDr. M. Zejda, Ph.D., RNDr. J. Jan{\'i}k, Ph.D. from the
Department of Theoretical Physics and Astrophysics, Masaryk University, Brno, Czechia, 
and Waldemar Ogloza from Pedagogical University of Cracow, Poland, for helping us with observations. 
\end{acknowledgements}

% for the bibliography, at the end
%\bibliographystyle{caosp310} % style aa.bst
\bibliography{refs.bib}

\end{document}